\documentclass[12pt]{article}

\usepackage{amsmath}
\usepackage{amsfonts,dsfont,mathrsfs}
\usepackage{amssymb}
\usepackage{graphics,graphicx,tikz,tensor}
\usetikzlibrary{arrows,decorations.pathmorphing,patterns}
\usepackage[linktocpage]{hyperref}
\hypersetup{
	colorlinks=true,
	linkcolor=blue,
	filecolor=magenta,      
	urlcolor=blue,
	citecolor=magenta
}
\numberwithin{equation}{section} 
\usepackage{caption,subcaption}
\usepackage{cancel}
\usepackage{tabu}
\usepackage{soul}
\usepackage{empheq}
\usepackage{color}

\usepackage[noadjust]{cite}

\begin{document}
	
\begin{titlepage}
	%\hfill \hbox{????}
	%\vskip 0.1cm
	%\hfill  \hbox{????}
	%\vskip 1.5cm
	\begin{flushright}
	\end{flushright}
	\vskip 1.0cm
	\begin{center}
		{\Large \bf Reconstructing the Gravitational Waveform\\ 
			from Its Probe Limit}
		\vskip 1.0cm {Carlo Heissenberg$^{\ast}$,
			Rodolfo Russo$^\dagger$} \\[0.7cm]
		{\it \small$^\ast$Institut de Physique Th\'eorique, CEA Saclay, CNRS, Universit\'e Paris-Saclay, \\
			F-91191, Gif-sur-Yvette Cedex, France}\\[0.5cm]
		{\it \small$^\dagger$School of Mathematical Sciences,
			Queen Mary University of London, \\
			Mile End Road, London, E1 4NS, United Kingdom.}
	\end{center}
	
	\vspace{5pt}
	
	\begin{abstract}
          Gravitational observables for binary systems exhibit a simple polynomial dependence on the masses $m_1$, $m_2$ of the two scattering objects when they are written in terms of the appropriate kinematic variables in the post-Minkowskian (PM) regime. We point out that this property, combined with particle interchange symmetry, allows one to reconstruct the leading and {\em subleading} PM waveforms from their probe limit, $m_1 \gg m_2$. 
          As an application, focusing on their re-expansion in the small-velocity or post-Newtonian (PN) regime, we calculate the probe-limit waveforms up to 10PN, and then exploit this observation to deduce from them the waveforms for generic masses in the center-of-mass frame up to 5PN. To this end, we employ both the amplitude-based waveform integrands and the tail formula. This combined approach simplifies substantially the PN expansion of the full subleading PM waveform. 
	\end{abstract}
\end{titlepage}

\tableofcontents

\section{Introduction}

In Newtonian mechanics, the dynamics of two particles interacting gravitationally is entirely determined by the probe limit, where one mass is much larger than the other, $m_1 \gg m_2$. This happens because, after subtracting  the free motion of the center of mass, the  relative motion coincides with that of a probe with effective mass $m_1 m_2/(m_1+m_2)$. In General Relativity (GR), this simple property does not hold and this makes the study of gravitational binaries a challenging problem for which no exact analytic solution is known. 

It is interesting to investigate at what order in the weak-field, or post-Minkowskian (PM), expansion one encounters the first contributions that cannot be determined by the probe limit. For instance, it turns out that all velocity-dependent corrections to the Newtonian deflection angle in a gravitational scattering are fully determined by the geodetic motion in the Schwarzschild solution to both leading and subleading PM order.\footnote{In this work we focus on the case of scalar particles and so neglect the effects due to spin.} This means that the scattering angle is determined by its probe limit up to 2PM.
Starting at 3PM order, the probe approximation is not sufficient to reconstruct the full scattering angle even in the conservative sector~\cite{Damour:2017zjx} and new dynamical information is needed. In the amplitude-based approach, this is obtained by taking the classical limit of the scattering of two minimally-coupled massive scalars and the conservative 3PM result was first derived from the corresponding $2\to 2$ amplitude at two loops~\cite{Bern:2019nnu,Bern:2019crd}.\footnote{At 3PM there are non-conservative, radiation-reaction contributions, which also go beyond the probe limit: in the PM regime they were first derived in ${\cal N}=8$ supergravity~\cite{DiVecchia:2020ymx} and then in pure GR~\cite{DiVecchia:2021bdo,Herrmann:2021tct} (see also~\cite{Damour:2020tta} for a derivation which does not use the amplitudes approach).} 

Additionally, the 2PM dynamics is entirely conservative and local, so it is possible to analytically continue the classical observables, such as the scattering angle, to the bound case~\cite{Kalin:2019rwq,Kalin:2019inp}, which implies that the 1PN (post-Newtonian) dynamics is determined by the probe limit also in the case of closed trajectories. In summary, as is well known, the first contributions to the conservative dynamics that go beyond the geodesic motion in the Schwarzschild metric appear at 3PM in the weak field expansion and at 2PN in the standard counting used when velocities are also small.

It is natural to ask if a similar pattern holds also in the radiative sector. In this paper we provide a positive answer to this question by showing that the PM  waveform produced in a binary scattering is fully determined by its probe limit both at the leading~\cite{Peters:1970mx,Kovacs:1978eu,Jakobsen:2021smu,Mougiakakos:2021ckm} and at   subleading~\cite{Brandhuber:2023hhy,Herderschee:2023fxh,Elkhidir:2023dco,Georgoudis:2023lgf,Caron-Huot:2023vxl,Bini:2023fiz,Georgoudis:2023eke,Georgoudis:2023ozp,Georgoudis:2024pdz,Bini:2024rsy} order. At leading order this was already noticed in~\cite{Kovacs:1978eu} and has a natural explanation in the amplitude based approach: at this order the waveform is determined by the tree-level $2\to 3$ amplitude with two minimally-coupled massive scalars in the incoming state and the same scalars plus a graviton in the final state. In the classical limit, the explicit mass dependence of the tree-level result appears just as an overall factor  when the kinematic dependence is expressed in terms of the four-velocities of the particles and the graviton momentum. The first subleading PM correction to the waveform was derived by computing the classical limit of the one-loop contribution to the same $2\to 3$ amplitude mentioned above~\cite{Brandhuber:2023hhy,Herderschee:2023fxh,Elkhidir:2023dco,Georgoudis:2023lgf} plus, as pointed out in~\cite{Caron-Huot:2023vxl}, some non-trivial cut contributions following from the KMOC approach~\cite{Kosower:2018adc}. See also \cite{Brunello:2025cot,Brunello:2025eso} for recent analytic results. We shall see that, exactly as it happens for NLO scattering angle, also the full NLO PM waveform can be derived from its probe limit.

A key role in our approach is played by the action of Lorentz transformations on the waveform. This is a well studied topic, see for instance~\cite{Cusin:2024git}, also because it can be useful phenomenologically when discussing the effects of the relative motion between the observer and the source. Here we use a Lorentz boost to connect the center-of-mass frame to the frame where one of the particle is initially at rest, which we shall refer to as rest frame for simplicity. Of course, in the full PM regime, this is a {\em finite} transformation, but it becomes easier to implement it explicitly when expanding also in the small velocity (PN) approximation. At leading PM order we show that an appropriate boost on the probe waveform yields the full center-of-mass waveform, as mentioned in~\cite{Kovacs:1978eu} (we check this explicitly up to 5PN order). At the first subleading PM order, the probe waveform does not directly capture the full result in rest mass frame of the heavy particle. We have to first find the complete rest-frame waveform and then perform the appropriate transformation to the center of mass frame. Exactly as it happens in the case of the deflection angle, the first step can be implemented by using the interchange symmetry between the two particles. In order to do so in the case of the waveform, one needs to take into account its dependence on the translation frame and its tensorial nature. We discuss these points in some detail and provide an algorithm that applies for generic velocity and can be easily implemented explicitly order by order in the velocity expansion. Again, we explicitly apply it to obtain the center-of-mass waveform up to 5PN, finding agreement with expressions that are available in the literature up to 3PN \cite{Georgoudis:2024pdz,Bini:2024rsy}.

An interesting feature of the probe limit of the waveform is that it dramatically simplifies the analysis of the so-called Compton contributions, which encode the rescattering of gravitational waves against the curvature sourced by the binary. Evaluating the PN expansion of these terms from the full PM result is the most computationally intensive part of the analysis of~\cite{Bini:2023fiz,Georgoudis:2023eke,Georgoudis:2024pdz,Bini:2024rsy}. In the probe limit, their contribution is \textit{entirely} captured by the tail formula and all the non-linear redefinitions needed in the Multipolar post-Minkowskian approach (see~\cite{Bernard:2017bvn,Blanchet:2023bwj,Blanchet:2023sbv,Bini:2023fiz} and references therein) are taken care of by the procedure we use to reconstruct the center-of-mass waveform. Conversely, the PN terms that are not captured by the Compton cuts are more easily derived from the amplitude result. These contributions are divided in two classes, as already noticed~\cite{Georgoudis:2023eke}: a first one which is equal to the leading order PM waveform, times a simple velocity-dependent prefactor, and a second one which requires the knowledge of (the real part of) the $2\to 3$ amplitude. We use this amplitude-inspired reorganization of the subleading waveform to generate the probe result from which we then reconstruct the full waveform in the center-of-mass frame.

Naturally, the probe limit represents the leading order of the gravitational self-force expansion, in which calculations are organized in powers of the mass ratio of the binary and which has been also recently combined with amplitude-inspired methods \cite{Cheung:2023lnj,Kosmopoulos:2023bwc,Cheung:2024byb,Damgaard:2024fqj,Mougiakakos:2024nku,Mougiakakos:2024lif,Akpinar:2025huz} (see also \cite{Adamo:2022qci,Adamo:2023cfp,Fucito:2024wlg,Cristofoli:2025esy,Cipriani:2025ikx}). It will be interesting to further explore and generalize to higher orders the connection between the PM and self-force expansions leveraging the approach outlined in this work.

The paper is organized as follows. In Section~\ref{sec:recon}, after briefly recalling the structure of the conservatives impulse, we detail the steps that we take to calculate the PM-expanded waveform starting from its probe-limit. We then present the explicit application of this algorithm at leading and subleading PM orders, working up to 10PN accuracy in the rest frame and 5PN accuracy in the center-of-mass frame. The explicit results are collected in the ancillary files, while in the text we provide as a simple application new even-in-velocity corrections to the $\mathcal{O}(G^4)$ angular momentum loss, complementing the results of \cite{Bini:2022enm,Geralico:2025rof}. Section~\ref{sec:Compton} discusses the amplitude-based representation of the Compton cuts and its connection with the tail formula in the multipolar representation.
Appendix~\ref{app:Polarizations} collects our conventions on physical polarizations and reviews their transformation properties under Lorentz symmetry.

\section{Reconstructing the PM observables from the probe limit}
\label{sec:recon}

We start with a brief summary of the kinematics of the gravitational scattering of two massive scalar particles. The initial state is described by the momenta $p_i^\mu$ (which we take to be outgoing) of the two incident particles\footnote{Throughout this paper we use with the mostly plus signature for the metric.}
\begin{equation}\label{eq:p-v}
	p_1^\mu = -m_1 v_1^\mu\,,\qquad 	p_2^\mu = -m_2 v_2^\mu\,,\qquad
	v_1^2 = -1\,,\qquad v_2^2 = -1\,,
\end{equation}
where the $v_i^\mu$ are the covariant velocities. The relative Lorentz factor is given by
\begin{align}\label{eq:sigma}
	\sigma = -\frac{p_1\cdot p_2}{m_1 m_2} = - v_1\cdot v_2 = \sqrt{1+p_\infty^2}\;,
\end{align}
where we introduced also the PN expansion parameter $p_\infty$. For instance, in the frame where the heavy particle is in the origin and at rest, we have
\begin{equation}\label{eq:rest}
		v_1^\mu = 
		(
		1,0,0,0
		)\,,
		\qquad
		v_2^\mu = (
		\sigma, 0,-p_\infty,
		0
		)
\end{equation}
and the Lorentz boost connecting the center-of-mass and the rest frames is given in~\eqref{eq:boosty}. Since we are interested in a scattering process, the final state includes again two massive particles, which in the PM approximation are slightly deflected with respect to the original direction of motion, plus radiation. As usual, we use the Newton constant $G$ as a formal PM expansion parameter and up to order $G^2$ (2PM) the motion of the massive particles is purely conservative. Thus by indicating the final momenta with $p_{3,4}$, we have 
\begin{equation}
p^\mu_1 +p^\mu_4 = {Q^\mu} +{\mathcal O}(G^3)\,, \qquad p^\mu_2 +p^\mu_3= -{Q^\mu}+{\mathcal O}(G^3)\,,
	\label{eq:barp}
\end{equation}
where $Q$ is the momentum exchange or impulse in the conservative approximation (the leading contribution to the energy carried away by gravitational waves is at 3PM and so it is hidden in the ${\mathcal O}(G^3)$ corrections).

A neat property of observables in the PM expansion is that, at each order, they depend in a simple polynomial way on the masses $m_1$, $m_2$ of the two objects taking part in the scattering. For instance, in the amplitude approach, the masses appear in a polynomial way in the vertices, by using~\eqref{eq:p-v}, or as an inverse power in the linearized propagators for the massive states. The classical limit ensures that only contributions scaling as $G^n m_1^{k+1} m_2^{n-k}$ wth $k=0,1,\ldots,n-1$ survive. In particular, this simple mass dependence was exploited in~\cite{Damour:2019lcq} to deduce that the 1PM and 2PM impulse for scatterings of objects with comparable mass are in fact entirely determined by the probe-limit setup. This perhaps surprising observation follows from the fact that the impulse, expressed as a function of the impact parameter and of the relative Lorentz factor $\sigma$, takes the form
\begin{equation}\label{eq:Q1Q2}
	Q_\text{1PM}^\mu = \hat{b}_e^\mu\,\frac{Gm_1m_2}{b_e}\,f_0(\sigma)\,,\qquad
	Q_\text{2PM}^\mu = \hat{b}_e^\mu\,\frac{G^2m_1m_2}{b_e^2}\,[m_1\,f_{m_1}(\sigma) + m_2\,f_{m_2}(\sigma)]\;.
\end{equation}
In the above,
\begin{equation}\label{eq:be}
	b_e^\mu = b^\mu  - \left(
	\frac{\sigma v_2^\mu-v_1^\mu}{2m_1 (\sigma^2-1)}
	-
	\frac{\sigma v_1^\mu-v_2^\mu}{2m_2 (\sigma^2-1)}
	\right)
	Q b + \mathcal{O}(G^2)
\end{equation}
is the {\em eikonal} impact parameter (with $\hat{b}_e^\mu=b_e^\mu/b_e$) which is orthogonal to 
\begin{equation}\label{eq:ptilde12}
		-\tilde{p}_1^\mu = p_1^\mu-Q^\mu/2\,,\qquad
		-\tilde{p}_2^\mu=p_2^\mu+Q^\mu/2
\end{equation}
rather than just $p_1^\mu$ and $p_2^\mu$, whereas the standard impact parameter $b^\mu$ obeys $p_{1,2}\cdot b=0$ and is directly related to the initial angular momentum in the center-of-mass frame, see~\cite{Cristofoli:2021jas,DiVecchia:2023frv,Georgoudis:2023eke} and references therein for more details. Notice that, in the center-of-mass frame, $b_e$ is rotated by $\Theta_c/2$ with respect to $b$, where $\Theta_c$ is the conservative scattering angle
\begin{equation}
  \label{eq:thcon}
  Q = 2 p \sin\frac{\Theta_c}{2}
\end{equation}
and $p$ is the spatial momentum. This difference will be relevant later on when discussing the subleading corrections to radiative observables. Going back to~\eqref{eq:Q1Q2}, the probe limit $m_1\gg m_2$ fixes both $f_0$ and $f_{m_1}$. However, by \eqref{eq:barp}, the complete conservative result should simply change sign when labels $1$ and $2$ are interchanged, and, since $b^\mu = b_1^\mu-b_2^\mu$, this requires $f_{m_1}(\sigma) = f_{m_2}(\sigma)$.
Thus, the first structure that cannot be fixed from the probe-limit calculation arises at 3PM \cite{Bern:2019nnu,Bern:2019crd}, since at this order the transverse part of the impulse is given by
\begin{equation}
	Q_\text{3PM}^\mu = \hat{b}_e^\mu\,\frac{G^3m_1m_2}{b_e^3} \left[m_1^2 \,f_{m_1^2}(\sigma) + m_1 m_2 \,f_{m_1 m_2}(\sigma)+ m_2^2\,f_{m_2^2}(\sigma)\right] ,
\end{equation}
so that, although $f_{m_1^2}(\sigma) = f_{m_2^2}(\sigma)$ by symmetry, $f_{m_1 m_2}(\sigma)$ is independent and indeed starts contributing in GR at second post-Newtonian (PN) order for small velocities \cite{Damour:2019lcq}.

The main objective of this work is to point out that a similar mechanism is at play in the case of the gravitational waveform, and to clarify how this observation can help simplify its calculation by first focusing on the probe limit, and then using particle-interchange symmetry. We will show that, in analogy with the case of the impulse summarized above, the probe-limit waveform is indeed enough to reconstruct the leading (tree-level) and subleading (one-loop) PM waveforms.

\subsection{Reconstructing the waveform: generalities}
\label{ssec:wav_g}

The central object in our analysis is the spectral waveform $\tilde{W}^{\mu\nu}(k)$, which determines the dynamical metric fluctuation $h_{\mu\nu} = g_{\mu\nu} - \eta_{\mu\nu}$ produced by the gravitational scattering via
\begin{equation}\label{eq:metricfluctuation}
	h_{\mu\nu} \sim \frac{4G}{\kappa r} \int_0^{\infty} \tilde{W}_{\mu\nu}(\omega n)\,e^{-i\omega U}\,\frac{d\omega}{2\pi} + \text{c.c.}\,,
\end{equation}
where $r$ is the (large) distance from the asymptotic detector, while $U$ and $n^\mu$ characterize its retarded time and angular direction (see e.g.~\eqref{eq:parametrization}), and  $\kappa = \sqrt{8\pi G}$. The dynamical information is contained in the two transverse traceless polarizations of gravitational waves, $\varepsilon_{A}^{\mu\nu} \tilde{W}_{\mu\nu}$ with $A=\pm$, while the covariant quantity $\tilde{W}^{\mu\nu}$, which is more natural in the amplitude approach, obeys $k_\mu \tilde{W}^{\mu\nu}=0$. The asymptotic metric also contains static and longitudinal components due to the Coulombic potential, which we do not consider here.

We start by adopting the symbol
\begin{equation}
	\mathcal{\tilde{W}}^{\mu\nu} = \mathcal{\tilde{W}}^{\mu\nu}(m_1, m_2, v_1, v_2, b, k) = \frac{\tilde{W}^{\mu\nu}}{\kappa m_1 m_2}
\end{equation}
for the mass-rescaled waveform in the translation frame where the origin is in the midpoint of the two particles in the transverse plane,
\begin{equation}\label{eq:defmidpoint}
	b_1^\mu = -b_2^\mu = \frac{1}{2}\,b^\mu\,.
\end{equation}
A convenient feature of this frame is that its definition respects particle-interchange symmetry $1\leftrightarrow 2$ and does not involve the masses. 
Order by order in the PM expansion,
\begin{equation}
	\mathcal{\tilde{W}}^{\mu\nu} = \mathcal{\tilde{W}}_0^{\mu\nu} + \mathcal{\tilde{W}}_{1}^{\mu\nu} + \mathcal{\tilde{W}}_{2}^{\mu\nu} + \cdots \,,
\end{equation}
the $\mathcal{\tilde{W}}_L^{\mu\nu}$ are polynomials in the masses $m_1$, $m_2$,
\begin{subequations}
	\begin{align}
		\label{eq:W0}
		\mathcal{\tilde{W}}_{0}^{\mu\nu} &= \mathcal{\tilde{W}}_{0}^{\mu\nu}(v_1,v_2,b,k)\,,\\
		\label{eq:W1}
		\mathcal{\tilde{W}}_{1}^{\mu\nu} &= m_1\,\mathcal{\tilde{W}}_{m_1}^{\mu\nu}(v_1,v_2,b,k) + m_2\,\mathcal{\tilde{W}}_{m_2}^{\mu\nu}(v_1,v_2,b,k)\,,\\
		\label{eq:W2}
		\mathcal{\tilde{W}}_{2}^{\mu\nu} &= m_1^2\,\mathcal{\tilde{W}}_{m_1^2}^{\mu\nu}(v_1,v_2,b,k) + m_1 m_2\,\mathcal{\tilde{W}}_{m_1 m_2}^{\mu\nu}(v_1,v_2,b,k)+m_2^2\,\mathcal{\tilde{W}}_{m_2^2}^{\mu\nu}(v_1,v_2,b,k) \,,
	\end{align}
\end{subequations}
where the tensors $\mathcal{\tilde{W}}_{m_1^\alpha m_2^\beta}^{\mu\nu}$ are independent of  $m_1$, $m_2$. 
This property has been verified by explicit calculations for tree level and one loop, and naturally follows from the amplitude representation of the waveform integrand. Particle-interchange symmetry requires that
\begin{equation}\label{eq:symmetryW}
	\mathcal{\tilde{W}}^{\mu\nu}_{m_1^\alpha m_2^\beta}(v_1, v_2, b, k)
	=
	\mathcal{\tilde{W}}^{\mu\nu}_{m_1^\beta m_2^\alpha}(v_2, v_1, -b, k)\,.
\end{equation}

In order to isolate the physical degrees of freedom, it is convenient to further decompose these building blocks by means of two transverse tensor structures with definite behavior under $1\leftrightarrow2$. We define
\begin{equation}\label{eq:xi1xi2}
	\xi_1^\mu = b\cdot k \, v_1^\mu + \omega_1 \, b^\mu \,,\qquad
	\xi_2^\mu =  - b\cdot k\, v_2^\mu - \omega_2\, b^\mu\,,
\end{equation}
with
\begin{equation}
	\label{eq:omega1omega2}
	\omega_1 = - v_1\cdot k\,,\qquad
	\omega_2 = -v_2\cdot k\,,
\end{equation}
so that $\xi_{1,2}^\mu = (\xi_{2,1}^\mu)_{1\leftrightarrow 2}$, and construct (the inverse powers of $p_\infty$ are introduced for convenience in the PN limit)
\begin{equation}\label{eq:xipmdef}
	\xi_+^{\mu\nu} = \frac{1}{p_\infty^2}\left(
	\xi_1^\mu \xi_2^\nu + \xi_2^\mu \xi_1^\nu
	\right),
	\qquad
	\xi_-^{\mu\nu} = \frac{1}{p_\infty^3}\left(
	\xi_1^\mu \xi_1^\nu - \xi_2^\mu \xi_2^\nu
	\right),
\end{equation}
which satisfy
\begin{equation}\label{eq:pmxipm}
	\xi_{\pm}^{\mu\nu} =\pm \left(\xi_{\pm}^{\mu\nu}\right)_{1\leftrightarrow2}\,.
\end{equation}
We can then adopt the decomposition
\begin{equation}\label{eq:WA}
	\mathcal{\tilde{W}}^{\mu\nu}_{m_1^\alpha m_2^\beta}(v_1, v_2, b, k)
	=
	\xi_A^{\mu\nu}
	\mathcal{\tilde{W}}_{m_1^\alpha m_2^\beta}^A(\sigma,\omega_1,\omega_2,b\cdot k)\,,
\end{equation}
where $A$ runs over the two values $A=\pm$. Of course this decomposition is valid up to gauge transformations, such as terms proportional to $k^\mu$ or $k^\nu$, which we can ignore for our purposes since we focus on physical observables by contracting with transverse traceless polarizations $\varepsilon_{A}^{\mu\nu}$. Denoting
\begin{equation}\label{eq:defM}
	\tilde{\mathcal{W}}_{m_1^\alpha m_2^\beta\,A}
	=
	\varepsilon_{A\mu\nu} \tilde{\mathcal{W}}^{\mu\nu}_{m_1^\alpha m_2^\beta}
	\,,
	\qquad
	M\indices{_{AB}} = \varepsilon_{A}^{\mu\nu} \xi^{\phantom{\mu}}_{B \mu\nu}\,,
\end{equation}
then $\mathcal{\tilde{W}}_{m_1^\alpha m_2^\beta} ^{\pm}$ can be obtained from
\begin{equation}
	\mathcal{\tilde{W}}^{A}_{m_1^\alpha m_2^\beta} = (M^{-1})^{AB} \,	\tilde{\mathcal{W}}_{m_1^\alpha m_2^\beta\,B}\,.
\end{equation}
Notice that the matrix~\eqref{eq:defM} is not symmetric in general, so one needs to be careful and lower or raise the indices by acting with $M$ or $M^{-1}$ from the left. The definitions \eqref{eq:xipmdef} were chosen so that $M_{AB}$ and its inverse both scale  as $p_\infty^0$ to leading order in the PN limit.
Thanks to the symmetry relation \eqref{eq:symmetryW}, and owing to the property \eqref{eq:pmxipm}, by~\eqref{eq:WA} we find that
\begin{equation}\label{eq:symmetryf}
	\mathcal{\tilde{W}}^{\pm}_{m_1^{\beta}m_2^{\alpha}}(\sigma,\omega_1,\omega_2,b\cdot k)
	=
	\pm \mathcal{\tilde{W}}^{\pm}_{m_1^{\alpha}m_2^{\beta}}(\sigma,\omega_2,\omega_1,-b\cdot k)\,.
\end{equation}

The above properties allow one to reconstruct $\mathcal{\tilde{W}}_{m_1^\alpha}^{\mu\nu}$ and $\mathcal{\tilde{W}}_{m_2^\alpha}^{\mu\nu}$ (for any $\alpha=0,1,\ldots$) from the waveform $\tilde{w}^{\mu\nu}$ in probe limit. This is the mass-rescaled waveform calculated in the limit $m_2\to0$ in the frame where $m_1$ is initially at rest in the origin. Its relation with $\mathcal{\tilde{W}}^{\mu\nu}$ introduced above is given by
\begin{equation}\label{eq:probewaveform}
	\tilde{w}^{\mu\nu} = e^{\frac{i}{2}\,b\cdot k} \,\mathcal{\tilde{W}}^{\mu\nu}\big|_{m_2\to0}\,,
\end{equation}
where the phase takes into account the translation by $a^\mu = -\frac{1}{2}\,b^\mu$ that is needed to move from the frame where $b_1^\mu = \frac{1}{2}\,b^\mu$ to the one where $b_1^\mu=0$. From \eqref{eq:probewaveform}, which can be recast as
\begin{equation}
	\tilde{w}^{\mu\nu}
	=
	\tilde{w}_0^{\mu\nu}
	+
	m_1\, \tilde{w}_{m_1}^{\mu\nu}
	+
	m_1^2\, \tilde{w}_{m_1^2}^{\mu\nu}
	+\cdots
	=
	e^{\frac{i}{2}\,b\cdot k} 
	\left(
	\mathcal{\tilde{W}}_0^{\mu\nu}
	+
	m_1\, \mathcal{\tilde{W}}_{m_1}^{\mu\nu}
	+
	m_1^2\, \mathcal{\tilde{W}}_{m_1^2}^{\mu\nu}
	+\cdots
	\right),
\end{equation}
it is already conceptually clear that the probe waveform fixes all $\mathcal{\tilde{W}}_{m_1^\alpha m_2^0}^{\mu\nu}$, from which, using \eqref{eq:symmetryW} or \eqref{eq:symmetryf}, we can reconstruct $\mathcal{\tilde{W}}_{m_1^0m_2^\alpha }^{\mu\nu}$ by symmetry. Instead, the terms $\mathcal{\tilde{W}}_{m_1^\alpha m_2^\beta }^{\mu\nu}$ with $\alpha,\beta\neq0$ go beyond the probe limit and require different methods. See for instance Ref.~\cite{Georgoudis:2025vkk} for the calculation of \eqref{eq:W2} in the soft limit $k\to0$, which is dominated by linear and nonlinear memory and  includes contributions of the kind $\mathcal{\tilde{W}}_{m_1 m_2 }^{\mu\nu}$.

At the practical level, suppose that we calculated the two physical polarizations\footnote{We \emph{do not} include a complex conjugate in the polarization here.} 
\begin{equation}\label{eq:wA}
 \tilde{w}_{A}(\omega,\theta,\phi)
 =
 \varepsilon_{A}^{\mu\nu} \tilde{w}_{\mu\nu}
\end{equation} 
as functions of $\sigma$, $b$, $\omega$, $\theta$ and $\phi$ (we omit $\sigma$ and $b$ from the list of arguments because they are spectators in what follows). Here $\omega$, $\theta$, $\phi$ are obtained from the invariants $\omega_1$, $\omega_2$, $b\cdot k$ by evaluating them in the rest frame \eqref{eq:rest} and adopting the standard parametrization \eqref{eq:parametrization} for $k^\mu$ in that frame. Explicitly,
\begin{equation}
		\omega_1 = \omega \,,\qquad
		\omega_2 =\omega(\sigma + p_\infty\,\sin\theta\,\sin\phi )\,,\qquad
		b\cdot k = \omega b\, \sin\theta \, \cos\phi\,,
\end{equation}
which can be inverted as follows\footnote{We use the two-argument variant of the arctangent, $-\pi < \arctan(x,y) < \pi$, which takes into account which quadrant the point $(x,y)$ is in.}
\begin{equation}
	\omega = \omega_1\,,\quad
	\theta = \arcsin\sqrt{\frac{ (\omega_2-\sigma \omega_1)^2}{\omega_1^2(\sigma^2-1)}+\frac{(b\cdot k)^2}{b^2\omega_1^2}}\,,\quad
	\phi = \arctan\left(
	\frac{b\cdot k}{b \omega_1}
	,
	\frac{\omega_2-\sigma\omega_1}{\omega_1\sqrt{\sigma^2-1}}
	\right).
\end{equation}
From these relations we can obtain a mapping
\begin{equation}
  \label{eq:mapping}
  (\omega,\theta,\phi)\mapsto F(\omega,\theta,\phi)\,,
\end{equation}
which performs the symmetry transformation $(\omega_1,\omega_2,b\cdot k)\mapsto (\omega_2,\omega_1,-b\cdot k)$ at the level of $(\omega,\theta,\phi)$. From the building blocks $\tilde{w}_{m_1^\alpha A}=\varepsilon_{A\mu\nu}\tilde{w}_{m_1^{\alpha}}^{\mu\nu}$, we can calculate
\begin{equation}
	\mathcal{\tilde{W}}_{m_1^\alpha}^{A} = e^{-\frac{i}{2}\,b\cdot k} (M^{-1})\indices{^{A B}}\,\tilde{w}_{m_1^\alpha B}
\end{equation}
as functions of $\omega$, $\theta$, $\phi$. Then, we can reconstruct $\tilde{\mathcal{W}}_{m_2^\alpha}^{A}$, again as explicit functions of $\omega$, $\theta$, $\phi$, by performing the transformation $F$ and taking into account the appropriate sign in \eqref{eq:symmetryf}. To summarize, we can reconstruct both 
\begin{equation}\label{eq:Wm1}
  \begin{aligned}
	\mathcal{\tilde{W}}_{m_1^\alpha A} 
	& = \varepsilon_{A}^{\mu\nu} \xi_{+\mu\nu}
	\left(
	e^{-\frac{i}{2}\,b\cdot k} (M^{-1})\indices{^{+B}}\,\tilde{w}_{m_1^\alpha B}
	\right)
	+
	\varepsilon_{A}^{\mu\nu} \xi_{-\mu\nu}
	\left(
	e^{-\frac{i}{2}\,b\cdot k} (M^{-1})\indices{^{-B}}\,\tilde{w}_{m_1^\alpha B}
	\right)
    \\ & =  e^{-\frac{i}{2}\,b\cdot k}\, \tilde{w}_{m_1^\alpha A}\,,
    \end{aligned}
\end{equation}
where the second line follows by using~\eqref{eq:defM}, and\footnote{Since the quantities in the round parentheses in Eqs.~\eqref{eq:Wm1} and \eqref{eq:Wm2} are $\tilde{\mathcal{W}}^{\pm}_{m_1^\alpha}$, we note that the latter reduces to the former for $\alpha=0$ by the property \eqref{eq:symmetryf}.}
\begin{equation}\label{eq:Wm2}
	\mathcal{\tilde{W}}_{m_2^\alpha A} 
	=
	\varepsilon_{A}^{\mu\nu} \xi_{+\mu\nu}\,
	F\!\left(
	e^{-\frac{i}{2}\,b\cdot k} (M^{-1})\indices{^{+B}}\,\tilde{w}_{m_1^\alpha B}
	\right)%_F
	-
	\varepsilon_{A}^{\mu\nu} \xi_{-\mu\nu}\,
	F\!\left(
	e^{-\frac{i}{2}\,b\cdot k} (M^{-1})\indices{^{-B}}\,\tilde{w}_{m_1^\alpha B}
	\right),
\end{equation}
where one needs to use the mapping~\eqref{eq:mapping}. The waveform obtained in this way is evaluated in the frame where $b_1^\mu=\frac{1}{2}\,b^\mu$ and in the rest frame \eqref{eq:rest}. But we can transform it to any desired frame by multiplying by appropriate phases and boosting using \eqref{eq:finalspins}. To move to the frame where the center of mass is in the origin and at rest we need to multiply by
\begin{equation}
    e^{\frac{i}{2}\, b\cdot k} e^{-i\,\frac{E_2}{E}\,b\cdot k}  = e^{ \frac{i}{2} \frac{E_1-E_2}{E}\,b\cdot k}
\end{equation}
with
\begin{equation}
    	E_1 = \frac{m_1(m_1+\sigma m_2)}{\sqrt{m_1^2+m_2^2+2m_1m_2 \sigma}}\,,\qquad
	E_2 = \frac{m_2(m_2+\sigma m_1)}{\sqrt{m_1^2+m_2^2+2m_1m_2 \sigma}}\,
\end{equation}
and perform the boost $B_y$ in \eqref{eq:boosty}, for which the little group phase $\Theta$ appearing in \eqref{eq:finalspins} is given by \eqref{eq:phaseboosty}. In conclusion,
\begin{equation}\label{eq:WCM}
	\mathcal{\tilde{W}}_{\pm}^\text{CM} =  e^{ \frac{i}{2} \frac{E_1-E_2}{E}\,b\cdot k} e^{\mp 2i \Theta}\,[\mathcal{\tilde{W}}_{\pm}]_{B_y}\,,
\end{equation}
where the subscript $B_y$ means applying the boost \eqref{eq:boosty} by transforming $\mathcal{\tilde{W}}_{\pm}$ accordingly.

\subsection{Tree level}

The leading PM waveform is expressed in terms of the tree-level amplitude $\mathcal{A}_0$ with one graviton emission,
\begin{equation}\label{eq:W0A0tree}
	\tilde{W}^{\mu\nu}_0
	=
	\kappa m_1 m_2 \tilde{\mathcal{W}}_0^{\mu\nu} = \tilde{\mathcal{A}}^{\mu\nu}_0
\end{equation}
where $\tilde{\mathcal{A}}_0$ is related to $\mathcal{A}_0$ by the inelastic Fourier transform, which for a generic function $f^{\mu\nu}(q_1,k)$ takes the form
\begin{equation}\label{eq:tildeFT}
	\tilde{f}^{\mu\nu}(k) = e^{-ib_2\cdot k} \int \frac{d^Dq_1}{(2\pi)^D}\,e^{ib\cdot q_1} 2\pi\delta(2p_1\cdot q_1)\,2\pi\delta(2p_2\cdot(q_1+k))\,f^{\mu\nu}(q_1,k)\,.
\end{equation}
Although $\tilde{W}^{\mu\nu}_0$ and $\tilde{\mathcal{A}}_0^{\mu\nu}$ are proportional to each other via \eqref{eq:W0A0tree}, as we shall see in the next section the relation between $\tilde{W}^{\mu\nu}$ and $\tilde{\mathcal{A}}^{\mu\nu}$ will be instead more involved at one loop (see \eqref{eq:W1full} below).
In this case, $\mathcal{W}_0$ is mass-independent when written in terms of $v_1$, $v_2$, $b$ and $k$. In other words, there is only one structure, which is fixed by the probe-limit calculation. Therefore, we can expand in the PN limit the amplitude $\mathcal{A}_0$ contracted with physical polarizations in the rest frame of particle 1, obtaining $\tilde{w}_{0\pm}$. Then, thanks to~\eqref{eq:Wm1} for $\alpha=0$, we can directly perform the appropriate boost and translation to calculate the center-of-mass waveform polarizations $W_{0\pm}^\text{CM}$ \eqref{eq:WCM}. We present these results in the ancillary file including up to 10PN corrections for the rest-frame waveform, that is N$^{20}$LO in the velocity, and 5PN (that is N$^{10}$LO in velocity) for the center-of-mass waveform.\footnote{We emphasize that this is merely a technical simplification. The 10PN rest-frame waveform, combined with the appropriately expanded boost, would  reproduce che 10PN center-of-mass frame waveform.}

At the practical level, rather than with $\theta$ and $\phi$, it can be convenient to work with the variables 
\begin{equation}\label{eq:tildeytildez}
	\tilde{y} = e^{i\theta}\,,\qquad \tilde{z} = e^{i\phi}\,.
\end{equation}
We find that $\tilde{w}_{0\pm}$ and $\tilde{\mathcal{W}}_{0\pm}^\text{PM}$ are real rational functions of $\tilde{y}$, $\tilde{z}$ at each PN order. Moreover, $+$ and $-$ get \emph{interchanged} when sending 
\begin{equation}\label{eq:plusminus}
	\tilde{y}\mapsto -\tilde{y}\,,\qquad 
	\tilde{z}\mapsto -\tilde{z}\,.
\end{equation}
This is actually a general property of the (spinless) waveform when the scattering plane is aligned with the $xy$ plane and the standard parametrization \eqref{eq:parametrization} and polarizations \eqref{eq:defeps1} are adopted. Indeed, recalling that the vectors $v_1$, $v_2$, $b$ have no component along the $z$ axis in this setup, for any vector $\xi$ chosen among $v_1$, $v_2$, $b$ one finds that under \eqref{eq:plusminus},
\begin{equation}
	\xi \cdot k \mapsto \xi \cdot k\,,\qquad
	\xi \cdot \varepsilon_{\pm} \mapsto  \xi \cdot \varepsilon_{\mp}\,,
\end{equation}
while the conditions $k\cdot \varepsilon_{\pm}=0=\varepsilon_{\pm}^2$ and $\varepsilon_+\cdot\varepsilon_-=1$ are preserved.
For this reason, in the ancillary file we present the results for the $-$ polarization as a function of $\tilde{y}$, $\tilde{z}$ and 
\begin{equation}
	u = \frac{\omega b}{p_\infty}\,,
\end{equation}
while the $+$ polarization can be easily obtained by applying \eqref{eq:plusminus}.

\subsection{One loop}

Moving to one loop, we recall that, at this order, the waveform kernel in the frame directly related to the initial momenta involves three different contributions: the real part of the one-loop $2\to3$ amplitude, $\tilde{\mathcal{B}}_1$, the so-called Compton or rescattering cuts (see Section~\ref{sec:Compton} below for their explicit expression), $\frac{i}{2}\,\tilde{c}_1$, $\frac{i}{2}\,\tilde{c}_2$, and the classical $s$-channel cut discussed in~\cite{Caron-Huot:2023vxl}. As shown in~\cite{Georgoudis:2023eke}, the latter contribution can be reabsorbed up to an overall time translation by switching to the eikonal frame, that is, by replacing, in the tree-level waveform, $b^\mu$ with $b_e^\mu$ and $p_{1,2}^\mu$ with $-\tilde{p}_{1,2}^\mu$, see Eqs.~\eqref{eq:be} and~\eqref{eq:ptilde12}. In the center-of-mass frame, this means rotating the scattering plane by $\Theta_c/2$,
\begin{equation}
  \label{eq:rotfe}
  \hat{b}^i_e = \cos\!\frac{\Theta_c}{2}\, b^i + \sin\!\frac{\Theta_c}{2} \, \hat{p}_{\rm cm}^i\,, \quad \hat{e}^i = -\sin\!\frac{\Theta_c}{2}\, \hat{b}^i + \cos\!\frac{\Theta_c}{2}\,\hat{p}_{\rm cm}^i\,,
\end{equation}
where $\hat{p}_{\rm cm}$ is the unit vector along the center-of-mass initial momentum of the particle $1$ and the index $i$ runs over the spatial components. Thus, we can drop the contribution of the $s$-channel cut and focus on the other two one loop contributions,  provided we interpret the tree-level result of the previous section as written in the eikonal frame. 

Let us start the discussion of these terms by recalling that $\mathcal{B}_1$ can be further decomposed as follows 
\begin{equation}
	\mathcal{B}_{1}^{\mu\nu} = \mathcal{B}_{1O}^{\mu\nu} +  \mathcal{B}_{1E}^{\mu\nu}
\end{equation}
into \emph{odd} and \emph{even} parts under the operation $q_{1,2}\mapsto -q_{1,2}$, $k\mapsto -k$.
The odd part $\mathcal{B}_{1O}$ is particularly simple and is determined by radiation-reaction in terms of the tree-level waveform discussed in the previous section:
\begin{equation}
	\mathcal{B}_{1O}^{\mu\nu} = \mathcal{B}_{1O}^{(i)\mu\nu} + \mathcal{B}_{1O}^{(h)\mu\nu}
\end{equation}
with
\begin{equation}\label{eq:B1OiB1Oh}
	\mathcal{B}_{1O}^{(i)\mu\nu} = - \pi G E \Omega\,\frac{\sigma(2\sigma^2-3)}{2(\sigma^2-1)^{3/2}}\, \mathcal{A}^{\mu\nu}_0\,,
	\qquad
	\mathcal{B}_{1O}^{(h)\mu\nu} = \pi G E \Omega\, \mathcal{A}^{\mu\nu}_0\,.
\end{equation}
Here we have introduced the total energy of the system $E$ and frequency $\Omega$ as measured in the center-of-mass frame
\begin{equation}\label{eq:EOmega}
	E = E_1+E_2= \sqrt{m_1^2+2 m_1 m_2 \sigma + m_2^2}\,,\qquad
	\Omega = \frac{m_1 \omega_1 + m_2 \omega_2}{E}\,.
\end{equation}
Instead, the even part $\mathcal{B}_{1E}$ is an instantaneous contribution that involves genuine one-loop data.
Finally, the Compton cuts involve a divergent part due to the long-range nature of gravity as $\epsilon\to0$ in $D=4-2\epsilon$, which can be isolated by letting\footnote{In \eqref{eq:irdiv}, $\mathcal{A}_0^{\mu\nu}$ is the tree-level amplitude in $D=4-2\epsilon$ dimension. This is relevant for the definition of the finite part ${c}_1^{(\text{reg})\mu\nu}$ \cite{Georgoudis:2024pdz,Bini:2024rsy}.}
\begin{equation}\label{eq:irdiv}
	\frac{i}{2}\,{c_1}^{\mu\nu}  = 2iGm_1\omega_1\left(
	-\frac{1}{2\epsilon} + \log\frac{\omega_1}{\mu}
	\right)
	{\mathcal{A}}_0^{\mu\nu} +\frac{i}{2}\, {c}_1^{(\text{reg})\mu\nu}\,.
\end{equation}
Being proportional to the tree-level amplitude, this IR divergence exponentiates and can be reabsorbed into a constant shift of the retarded time $U$ in Eq.~\eqref{eq:metricfluctuation} \cite{Goldberger:2009qd,Porto:2012as},
and the arbitrary scale $\mu$ reflects the freedom in performing further finite time translations. Therefore, recalling $\tilde{W}^{\mu\nu}_1
	=
	\kappa m_1 m_2 \mathcal{\tilde{W}}_1^{\mu\nu}$,
\begin{equation}\label{eq:W1full}
	\begin{split}
		\tilde{W}^{\mu\nu}_1
		&=
		\pi G (m_1\omega_1+m_2\omega_2) \left[-\frac{\sigma(2\sigma^2-3)}{2(\sigma^2-1)^{3/2}}+1\right] \tilde{\mathcal{A}}^{\mu\nu}_0
		\\
		&+  i G \left(
		m_1\omega_1\log\frac{\omega_1^2}{\mu^2}+
		m_2\omega_2\log\frac{\omega_2^2}{\mu^2}
		\right)
		\tilde{\mathcal{A}}_0^{\mu\nu}
		+
		\tilde{\mathcal{B}}_{1E}^{\mu\nu}
		+
		\tilde{\mathcal{C}}^{(\text{reg})\mu\nu}\,,
	\end{split}
\end{equation}
where
\begin{equation}
	\mathcal{C}^{(\text{reg})\mu\nu} =  
	\frac{i}{2}\, c_1^{(\text{reg})\mu\nu} 
	+
	\frac{i}{2}\,c_2^{(\text{reg})\mu\nu} \,.
\end{equation}
For later convenience, we can isolate the $\mu$-dependent terms by adding and subtracting $2iGE\Omega\,\log\Omega$ as follows,
\begin{equation}\label{eq:W1full2}
	\begin{split}
		\tilde{W}^{\mu\nu}_1
		&= 
		\pi G E\Omega\,\left[-\frac{\sigma(2\sigma^2-3)}{2(\sigma^2-1)^{3/2}}+1\right] \tilde{\mathcal{A}}^{\mu\nu}_0
		+  
		2iG E\Omega\, \log\frac{\Omega}{\mu} \,\tilde{\mathcal{A}}_0^{\mu\nu}
		\\
		&
		+
		i G \left(
		m_1\omega_1\log\frac{\omega_1^2}{\Omega^2}+
		m_2\omega_2\log\frac{\omega_2^2}{\Omega^2}
		\right)
		\tilde{\mathcal{A}}_0^{\mu\nu}
		+
		\tilde{\mathcal{B}}_{1E}^{\mu\nu}
		+
		\tilde{\mathcal{C}}^{(\text{reg})\mu\nu}\,,
	\end{split}
\end{equation}

There is a further freedom related to the BMS frame for the supertranslations, see~\cite{Veneziano:2022zwh} and references therein. The amplitude result~\eqref{eq:W1full} (or \eqref{eq:W1full2}) holds in the so-called canonical BMS frame where the initial shear vanishes, while the PN literature traditionally works in the intrinsic frame, where the initial shear is determined by the velocities of the incoming particles.\footnote{By defining appropriately the contribution of ``zero-frequency'' gravitons, it is possible to obtain the one loop waveform in the intrinsic BMS frame directly in the amplitude approach~\cite{Bini:2024rsy,Elkhidir:2024izo}.} Neglecting a static contribution proportional to $\delta(\Omega)$, the supertranslation that connects the two leads to the following waveform in the intrinsic BMS frame,
\begin{equation}\label{eq:STVV}
	\tilde{W}^{\mu\nu}\mapsto \tilde{W}^{\mu\nu} + i G \left(
	m_1\omega_1\log\frac{\omega_1^2}{\Omega^2}+
	m_2\omega_2\log\frac{\omega_2^2}{\Omega^2}
	\right) \tilde{\mathcal{A}}_0^{\mu\nu}
\end{equation}
i.e.~to \emph{doubling} the first term in the last line of \eqref{eq:W1full2}. For our purposes, it is more convenient to work in the canonical frame, i.e.~with the expression \eqref{eq:W1full}, because the procedure to reconstruct the waveform from the probe limit hinges on the fact that $\tilde{\mathcal{W}}_1^{\mu\nu}$ is manifestly linear in $m_1$, $m_2$ when expressed in terms of $v_1^\mu$, $v_2^\mu$, $b^\mu$, $k^\mu$ and of their invariant products, while \eqref{eq:STVV} introduces a nontrivial dependence  on the masses via~\eqref{eq:EOmega}.

In the following, we study the two ingredients in \eqref{eq:W1full} which are not related to the tree-level waveform, $\tilde{\mathcal{B}}_{1E}$ and $\tilde{\mathcal{C}}^{(\text{reg})}$. We focus on the $-$ polarization (but \eqref{eq:plusminus} allows us to easily switch between $+$ and $-$ polarization) and note that $\tilde{\mathcal{B}}_{1E}$ ($\tilde{\mathcal{C}}^{(\text{reg})}$) turns out to be purely real (imaginary) when written in terms of the variables \eqref{eq:tildeytildez}.  When expanding the full $\tilde{W}_1^{\mu\nu}$ given by \eqref{eq:W1full} the PN limit, one finds that the leading-order ``Newtonian'' $\mathcal{O}(p_\infty^{-3})$ contribution arises from the first line. Instead, the PN expansion of $\tilde{\mathcal{B}}_{1E}^{\mu\nu}$ begins with $p_\infty^{-1}$, and thus starts contributing at 1PN with respect to the Newtonian term, while the PN expansion of  $\tilde{\mathcal{C}}^{^\text{(reg)}\mu\nu}$ begins with $p_\infty^{0}$, and thus starts contributing at 1.5PN with respect to the Newtonian term. 

Starting from $\tilde{\mathcal{B}}_{1E}$, we find it convenient to perform the PN expansion in the probe limit directly of its amplitude-based representation \cite{Brandhuber:2023hhy,Herderschee:2023fxh,Elkhidir:2023dco,Georgoudis:2023lgf,Georgoudis:2023ozp}, by contracting with physical polarizations and aligning first the velocity along the $z$ axis. Then we perform the needed rotation to bring the velocity along the $y$ axis as in \eqref{eq:rest}, taking into account the little group phase \eqref{eq:lgphase-x}. In this way, we obtain the PN expansion of $\tilde{\mathcal{B}}_{1E}$ in the probe limit up to 9PN relative to its own leading order, this is, 10PN order compared to the Newtonian order in the waveform, since $\tilde{B}_{1E}$ starts at 1PN. We then employ the strategy detailed in Section~\ref{ssec:wav_g} to first reconstruct the other mass structure and then perform the needed translation and boost to obtain the full center-of-mass result. We present our explicit results for the latter up to relative 4PN (i.e.~5PN compared to Newtonian) order.

Finally we note that, since we can work to leading order in the mass ratio, the full $\tilde{\mathcal{C}}^{(\text{reg})}$ is actually fixed by the tail formula (see Section~\ref{sec:Compton} below) in terms of the multipole-expansion of the tree-level waveform $\tilde{\mathcal{A}}_0$,
\begin{equation}\label{eq:Ctail}
		\mathrm{U}_{\ell m}^{\mathcal{C}} 
		= -
		2iGm_1  \omega_1  \mathrm{U}_{\ell m}^\text{tree} \,\kappa_\ell\,,
\qquad
		\mathrm{V}_{\ell m}^{\mathcal{C}} 
		= -
		2iGm_1 \omega_1  \mathrm{V}_{\ell m}^\text{tree} \pi_\ell\,,
\end{equation}
where the numbers $\kappa_\ell$ and $\pi_\ell$ are given  by
\begin{equation}\label{eq:harmonicnumbers}
	\kappa_\ell
	= \frac{2\ell^2+5\ell+4}{\ell(\ell+1)(\ell+2)} + \sum_{k=1}^{\ell-2}\frac{1}{k}\,,
	\qquad
	\pi_\ell
	= \frac{\ell-1}{\ell(\ell+1)} + \sum_{k=1}^{\ell-1}\frac{1}{k}\,.
\end{equation}
We provide the probe-limit result for $\tilde{\mathcal{C}}^{(\text{reg})}$ up to relative 10PN order, which means 11.5PN order compared to the Newtonian waveform. We can then run the algorithm discussed above to reconstruct the other mass structure, and perform the translation and boost to go to the center-of-mass frame, where we present the result up to relative 3.5PN order (5PN compared to Newtonian level). We collect the explicit results discussed so far for $\tilde{\mathcal{A}}_0$, $\tilde{\mathcal{B}}_{1E}$, $\tilde{\mathcal{C}}^{(\text{reg})}$ in the  file \texttt{wf10PNprobe5PNcm.m} and briefly illustrate their use in the ancillary file \texttt{anc-wf10PNprobe5PNcm.nb}. 

We have checked that these results agree with the ones obtained in \cite{Georgoudis:2024pdz,Bini:2024ijq} up to 3PN in the center-of-mass frame.
As a further cross-check, we verified that integrating the corresponding fluxes (see e.g.~\cite{Heissenberg:2024umh}) reproduces the available results for the radiated energy \cite{Herrmann:2021lqe,Dlapa:2022lmu} and angular momentum \cite{Manohar:2022dea,Heissenberg:2025ocy}.

Focusing on the latter quantity, let us recall that the total angular momentum loss $J^{\alpha\beta}$ is given by the sum of the radiated angular momentum $\boldsymbol{J}^{\alpha\beta}$ and the static loss $\mathcal{J}^{\alpha\beta}$. 
Focusing on the $\mathcal{O}(G^4)$ contribution defined in the center-of-mass frame, we have $J_{\mathcal{O}(G^4)} = J_\text{1rad} + J_\text{2rad}$ where $J_\text{1rad}$ (resp. $J_\text{2rad}$) contains all even (odd) velocity corrections with respect to the leading ``Newtonian'' $\mathcal{O}(G^2p_\infty^2)$ contribution. The odd-in-velocity part $J_\text{2rad}$ was calculated for any velocity and in a generic frame in \cite{Heissenberg:2025ocy}, and our results agree with that reference.
For the even-in-velocity part, we find
\begin{equation}\label{eq:J1radCM}
	\begin{split}
		&J_{\text{1rad}}
		=
		\frac{G^4 M^5 \nu^2}{b^3 p_\infty^2}
		\Big[
		\frac{176}{5}
		+
		\left(
		\frac{8144}{105}-\frac{2984}{45}\nu
		\right)
		p_\infty^2
		+
		\left(
		\frac{722}{9}\nu^2-\frac{12182}{225}\nu-\frac{93664}{1575}
		\right)
		p_\infty^4
		\\
		&+
		\left(
		-\frac{1399 \nu ^3}{15}+\frac{71339 \nu ^2}{1575}+\frac{228317 \nu }{11025}-\frac{4955072}{121275}
		\right) p_\infty^6
		\\
		&+
		\left(
		\frac{7549 \nu ^4}{72}-\frac{56261 \nu ^3}{2100}-\frac{2675321 \nu ^2}{264600}+\frac{48723439 \nu }{970200}-\frac{29857664}{1576575}
		\right) p_\infty^8
		\\
		&+
		\left(
		-\frac{83027 \nu ^5}{720}+\frac{2519 \nu ^4}{1260}+\frac{29927 \nu ^3}{35280}-\frac{317449889 \nu ^2}{5821200}+\frac{1450030123 \nu }{99891792}+\frac{28280064}{1926925}
		\right)
		p_\infty^{10}
		\\
		&+\mathcal{O}(p_\infty^{12})
		\Big]
	\end{split}
\end{equation}
for generic $\nu$. We emphasize that, just like to waveform up to NLO PM order discussed above, also this $\mathcal{O}(G^4)$ angular momentum loss, which is obtained by suitably ``squaring'' the waveform, can be bootstrapped from the probe limit. Finally, to leading order in the probe limit $\nu\to0$, we obtain
\begin{equation}\label{eq:J1radProbe}
	\begin{split}
		J_{\text{1rad}}
		&=
		\frac{G^4 M^5 \nu^2}{b^3 p_\infty^2}
		\Big[
		\frac{176}{5}
		+
		\frac{8144 p_{\infty }^2}{105}
		-
		\frac{93664 p_{\infty }^4}{1575}
		-
		\frac{4955072 p_{\infty }^6}{121275}
		\\
		&
		-
		\frac{29857664 p_{\infty }^8}{1576575}
		+
		\frac{28280064 p_{\infty }^{10}}{1926925}
		-
		\frac{551268352 p_{\infty }^{12}}{32757725}
		+
		\frac{4799560603648 p_{\infty }^{14}}{218461268025}
		\\
		&
		-
		\frac{108904313454592 p_{\infty }^{16}}{3713841556425}
		+
		\frac{62716517482446848 p_{\infty }^{18}}{1622948760157725}
		-
		\frac{130331671789568 p_{\infty }^{20}}{2613444058225}
		\\
		&
		+
		\mathcal{O}(p_\infty^{22})
		\Big].
	\end{split}
\end{equation}
The first two lines of Eq.~\eqref{eq:J1radCM} agree with the results obtained in \cite{Bini:2022enm}, while the third and fourth lines provide two new PN orders for $J_\text{1rad}$. Similarly, the first two lines of Eq.~\eqref{eq:J1radProbe} match the result of Ref.~\cite{Geralico:2025rof}, while the third line provides three new PN orders for this quantity.

\section{Compton cuts revisited}
\label{sec:Compton}

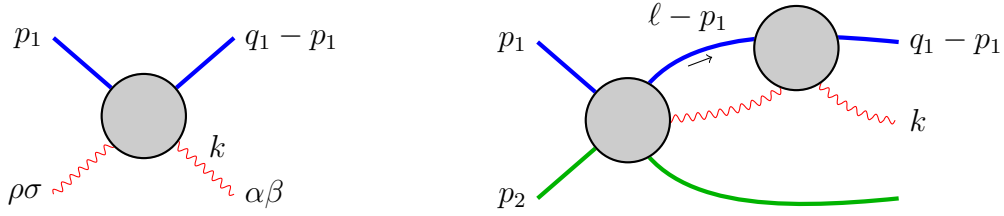
\begin{figure}
	\begin{center}
		\begin{tikzpicture}[scale=.8]
			\draw [red,decorate,decoration={coil,,aspect=0,segment length=1.5mm,amplitude=.5mm,pre length=0pt,post length=0pt}] (-3,-1.3) -- (-1.5,0);
			\draw [red,decorate,decoration={coil,,aspect=0,segment length=1.5mm,amplitude=.5mm,pre length=0pt,post length=0pt}] (0,-1.3) -- (-1.5,0);
			\draw [ultra thick, blue] (-3,1.3) -- (-1.5,0);
			\draw [ultra thick,blue] (0,1.3) -- (-1.5,0);
			\filldraw[black!20!white, thick] (-1.5,0) ellipse (.7 and .7);
			\draw[thick] (-1.5,0) ellipse (.7 and .7);
			%\draw[->] (-.2,-.2)--(.2,-.2);
			\node at (-3,-1.3)[left]{$\rho\sigma$};
			\node at (0,-1.3)[right]{$\alpha\beta$};
			\node at (-.6,-.5)[right]{$k$};
			\node at (-3,1.3)[left]{$p_1$};
			\node at (0,1.3)[right]{$q_1-p_1$};
		\end{tikzpicture}
		\hspace{45pt}
		\begin{tikzpicture}[scale=.8]
			\draw [ultra thick, green!70!black] (-3,-1.3) -- (-1.5,0);
			\draw [ultra thick, green!70!black] (-1.5,0) .. controls (-1,-1.5) and (1,-1.5) .. (3,-1.3);
			\draw[red,decorate,decoration={coil,,aspect=0,segment length=1.5mm,amplitude=.5mm,pre length=0pt,post length=0pt}] (-1.5,0)  .. controls (-1,0) and (1.5,0) .. (1.3,1.2);
			\draw[red,decorate,decoration={coil,,aspect=0,segment length=1.5mm,amplitude=.5mm,pre length=0pt,post length=0pt}] (2.94,0)  .. controls (2.5,0) and (1.7,.3) .. (1.3,1.2);
			%\draw[ultra thick] (-1.5,0)  .. controls (-1,-.9) and (1,-.9) .. (1.5,0);
			\draw [ultra thick, blue] (-3,1.3) -- (-1.5,0);
			\draw [ultra thick,blue] (-1.5,0) .. controls (-1,1.5) and (1,1.5) .. (3,1.3);
			\filldraw[black!20!white, thick] (-1.5,0) ellipse (.7 and .7);
			\draw[thick] (-1.5,0) ellipse (.7 and .7);
			\filldraw[black!20!white, thick] (1.3,1.2) ellipse (.7 and .7);
			\draw[->] (-.5,.9)  -- (-.1,1.05);
			\draw[thick] (1.3,1.2) ellipse (.7 and .7);
			\node at (-3,-1.3)[left]{$p_2$};
			\node at (3,0)[right]{$k$};
			\node at (-3,1.3)[left]{$p_1$};
			\node at (3,1.3)[right]{$q_1-p_1$};
			\node at (-.5,1.3)[above]{$\ell-p_1$};
		\end{tikzpicture}
	\end{center}
	\caption{On the left, a diagrammatic representation of the Compton amplitude given by Eq.~\eqref{eq:ComptonClassical}. On the right,  the Compton or rescattering cut constructed in Eq.~\eqref{eq:c1convolution}. \label{fig:Compton}}
\end{figure}

We consider the Compton cuts defined from the convolution of a $2\to3$ amplitude and a $2\to2$ Compton amplitude in the classical limit.\footnote{This is the analog of the limit in which Compton scattering reduces to Thomson scattering in electromagnetism. The distinguishing feature of gravity is the presence of the $1/q_1^2$ pole due to nonlinearities.}
See Figure~\ref{fig:Compton}.
The first ingredient can be found in \cite{Luna:2017dtq,DiVecchia:2021bdo}, while for the second one we start from the exactly transverse representation given in \cite{KoemansCollado:2019ggb,DiVecchia:2023frv} and take the classical limit $k^\mu \sim q_1^\mu \sim \mathcal{O}(\lambda)$ as $\lambda\to0$ for generic $m_1$ and  $p_1^\mu = - m_1 v_1^\mu$, obtaining\footnote{While the expression \eqref{eq:ComptonClassical} is not symmetric in $(\rho,\sigma)$, $(\alpha,\beta)$, it will be contracted with manifestly symmetric tensors later on.}
\begin{equation}\label{eq:ComptonClassical}
	\mathcal{A}^{(C)\rho\sigma\alpha\beta} (q_1,k) = \frac{32\pi G m^2_1\omega_1^2}{q_1^2}\,\Gamma^{\rho\alpha}(q_1,k)\,\Gamma^{\sigma\beta}(q_1,k)\,,
\end{equation}
with $\omega_1 = -v_1\cdot k$ 
\begin{equation}
	\Gamma^{\rho\alpha}(q_1,k) = \frac{1}{\omega_1}\left(k^\rho v_1^\alpha + k^\alpha v_1^\rho+q_1^\alpha v_1^\rho\right) - \frac{q_1^2}{2\omega_1^2}\,v_1^\rho v_1^\alpha\,+\eta^{\rho\alpha}\,.
\end{equation}
Note that the mass-shell conditions enforce\footnote{The second relation in \eqref{eq:massshellapprox} follows from $2 p_1\cdot q_1 = q_1^2$ approximated to leading order in $\lambda$.}
\begin{equation}\label{eq:massshellapprox}
	q_1^2+2q_1\cdot k=0\,,\qquad v_1\cdot q_1\approx 0\,,
\end{equation}
so that
\begin{equation}
	\Gamma^{\alpha\beta}(q_1,k)k_\alpha = 0\,,\qquad \Gamma^{\rho\sigma}(q_1,k)(k_\rho+q_{1\rho}) \approx 0
\end{equation}
in accordance with gauge invariance.
The Compton cut describing the rescattering of radiation against particle 1 then takes the form (see Figure~\ref{fig:Compton})
\begin{equation}\label{eq:c1convolution}
	c_1^{\alpha\beta}(q_1,k)
	=
	\int \frac{d^D\ell}{(2\pi)^D}\,
	2\pi\delta(2p_1\cdot \ell)\,2\pi\delta((k+q_1-\ell)^2)
	\mathcal{A}_0 (\ell,k-\ell+q_1) \mathcal{A}^{(C)\alpha\beta}(q_1-\ell,k)\,.
\end{equation}
Note that, provided $k^0>0$, since $p_1\cdot q_1\approx 0$ in the classical limit, then $k^0+q^0-\ell^0$ is positive on the support of the delta functions. We adopt a shorthand notation according to which indices are suppressed when contracted with the structure
\begin{equation}
	X Y = X^{\mu\nu}\left(\eta_{\mu\rho}\eta_{\nu\sigma}-\tfrac{1}{D-2}\,\eta_{\mu\nu}\eta_{\rho\sigma}\right) Y^{\rho\sigma}\,.
\end{equation}
Inserting the convolution \eqref{eq:c1convolution} into the Fourier transform \eqref{eq:tildeFT} and changing variable according to $q_1\mapsto q_1+\ell$, one finds
\begin{equation}\label{eq:c1tildecovariant}
	\tilde{c}_1^{\alpha\beta}(k) = \int \frac{d^Dq_1}{(2\pi)^D}\,e^{ib_1\cdot q_1}\,2\pi\delta(2p_1\cdot q_1)\,2\pi\delta((k+q_1)^2)\,\tilde{\mathcal{A}}_0(k+q_1)\mathcal{A}^{(C)\alpha\beta}(q_1,k)\,,
\end{equation}
where we have recognized the Fourier transform of the $2\to3$ amplitude in the  next-to-last factor.

Let us go to the frame where $b_1^\alpha=0$ and $v_1^\alpha=(1,0,0,0)$, and contract both sides with purely spatial polarizations vectors defined in this frame, 
\begin{equation}
	\varepsilon^\mu(\hat k)k_\mu =0=\varepsilon^\mu(\hat k)\varepsilon_\mu(\hat k)\,,\qquad
	v_1^\mu \varepsilon_{\mu}(\hat k) = 0\,,
\end{equation}
where $k^\alpha = \omega_1(1,\hat k^i)$ and $\varepsilon^\alpha = (0,\varepsilon^i)$.
Let us also express the $2\to3$ amplitude in transverse-traceless (TT) gauge
  \begin{equation}
    \label{eq:TTg}
    \begin{gathered}
      	\mathcal{A}^{\mu\nu}_{\text{TT}}(k) 
	= \Pi^{\mu\nu,\rho\sigma}(k) 
	\mathcal{A}_{\rho\sigma}(k)\,,\quad \mbox{with}
\\ 
      	 \Pi_{\mu\nu,\rho\sigma} = \frac{1}{2}\left(
	 \Pi_{\mu\rho} \Pi_{\nu\sigma} + \Pi_{\mu\sigma} \Pi_{\nu\rho}
	 -  \frac{2}{D-2} \Pi_{\mu\nu} \Pi_{\rho\sigma}
	 \right),
	 \quad
	 \Pi_{\mu\nu} = \eta_{\mu\nu} + \lambda_{\mu} k_{\nu} + \lambda_{\nu} k_{\mu}
    \end{gathered}
  \end{equation}
and $\lambda^2=0$, $\lambda\cdot k=-1$.
Then, we find the following simple expression result for the amplitude contraction,
\begin{equation}
	\tilde{\mathcal{A}}_{0\text{TT}}(k+q_1)\mathcal{A}^{(C)\alpha\beta}(q_1,k)\varepsilon_\alpha(\hat k) \varepsilon_\beta(\hat k)
	=
	32\pi G m_1^2\,\frac{\omega_1^2}{q_1^2}\, \tilde{\mathcal{A}}^{\alpha\beta}_{0\text{TT}}(k+q_1)\,\varepsilon_\alpha(\hat k) \varepsilon_\beta(\hat k)\,,
\end{equation}
and, letting
\begin{equation}
 \tilde{c}_1(k) = \tilde{c}_1^{\alpha\beta}(k) \varepsilon_\alpha(\hat k) \varepsilon_\beta(\hat k)\,,
\end{equation}
the $\delta$-functions in \eqref{eq:c1tildecovariant} can be solved to yield
\begin{equation}
	\frac{i}{2}\,\tilde{c}_1(k) = 2 i G m_1 \omega_1^{D-3} \int \frac{d\Omega_{D-2}(\hat n)}{2(2\pi)^{D-3}} 
	\,
	\frac{1}{1-\hat{n}\cdot \hat{k}}
	\,
	\tilde{\mathcal{A}}^{\alpha\beta}_{0\text{TT}}(\omega_1(1,\hat n))\,\varepsilon_\alpha(\hat k) \varepsilon_\beta(\hat k)\,.
\end{equation}
It is convenient to isolate the collinear divergence emerging as $\epsilon\to0$ with $D=4-2\epsilon$ by letting
\begin{equation}
	 \tilde{c}_1(k) =  \tilde{c}^\text{(div)}_1(k) +  \tilde{c}^\text{(reg)}_1(k)\,,
\end{equation}
where
\begin{subequations}
	\begin{align}
		\label{eq:c1div}
		\frac{i}{2}\,
		\tilde{c}^\text{(div)}_1
		&=
		2 i G m_1 \omega_1^{D-3}  \tilde{\mathcal{A}}^{\alpha\beta}_{0\text{TT}}(k)\,\varepsilon_\alpha(\hat k) \varepsilon_\beta(\hat k)
		\int \frac{d\Omega_{D-2}(\hat n)}{2(2\pi)^{D-3}}\,
		\frac{1}{1-\hat{n}\cdot \hat{k}}
		\,,
		\\ 
		\label{eq:c1reg}
		\frac{i}{2}\,
		\tilde{c}^\text{(reg)}_1
		&=
		2 i G m_1 \omega_1 \int \frac{d\Omega_{2}(\hat n)}{4\pi} 
		\,
		\frac{\varepsilon_\alpha(\hat k) \varepsilon_\beta(\hat k)}{1-\hat{n}\cdot \hat{k}}
		\left[
		\tilde{\mathcal{A}}^{\alpha\beta}_{0\text{TT}}(\omega_1(1,\hat n))
		-
		\tilde{\mathcal{A}}^{\alpha\beta}_{0\text{TT}}(\omega_1 (1,\hat k))
		\right].
	\end{align}
\end{subequations}
The integral in Eq.~\eqref{eq:c1div} evaluates to 
\begin{equation}\label{eq:resultdiv}
	\frac{i}{2}\,
	\tilde{c}^\text{(div)}_1
	=
	2i G m_1\omega_1
	\tilde{\mathcal{A}}^{\alpha\beta}_{0\text{TT}}(k)\,\varepsilon_\alpha(\hat k) \varepsilon_\beta(\hat k)
	\left(
	-\frac{1}{2\epsilon}+\log\frac{\omega_1}{\mu} + \mathcal{O}(\epsilon)
	\right),
\end{equation}
where $\mu$ is a running IR energy scale. In this way we see that the splitting into divergent and regularized part introduced here is the same as in the previous section, see Eq.~\eqref{eq:irdiv}. To discuss \eqref{eq:c1reg}, let us introduce the decomposition of the TT waveform in terms of symmetric trace-free (STF) multipoles (see e.g.~\cite{Blanchet:1985sp,Blanchet:2013haa} for more details), and switch to spatial indices for clarity,
\begin{equation}
	\tilde{\mathcal{A}}_{0\text{TT}}^{ij}(k) = \Pi^{ijab}(\hat k) \tilde{\mathcal{A}}^{ab}_{0\text{STF}}(k)\,,
\end{equation}
so we obtain
\begin{equation}
	\tilde{\mathcal{A}}^{ab}_{0\text{STF}}(k) = \sum_{\ell=0^\infty}\frac{1}{\ell!}\left[
	\hat k^{L-2}
	\mathrm{U}^\text{tree}_{ab L-2}
	- \frac{2\ell}{\ell+1}\,\hat k^{c L-2} \varepsilon_{cd(a} \mathrm{V}^\text{tree}_{b)dL-2}
	\right].
\end{equation}
Then,
\begin{equation}\label{eq:c1regintermediate}
	\begin{split}
		\frac{i}{2}\,
		\tilde{c}^\text{(reg)}_1
		&=
		2 i G m_1 \omega_1 \int \frac{d\Omega_{2}(\hat n)}{4\pi} 
		\,
		\frac{\varepsilon^i(\hat k) \varepsilon^j(\hat k)}{1-\hat{n}\cdot \hat{k}}
		\left[
		\Pi^{ijab}(\hat n)
		-
		\Pi^{ijab}(\hat k)
		\right]
		\tilde{\mathcal{A}}^{ab}_{0\text{STF}}(\omega_1(1,\hat n)) 
		\\
		&+
		2 i G m_1 \omega_1 \int \frac{d\Omega_{2}(\hat n)}{4\pi} 
		\,
		\frac{\varepsilon^a(\hat k) \varepsilon^b(\hat k)}{1-\hat{n}\cdot \hat{k}}
		\left[
		\tilde{\mathcal{A}}^{ab}_{0\text{STF}}(\omega_1(1,\hat n)) 
		-
		\tilde{\mathcal{A}}^{ab}_{0\text{STF}}(\omega_1 (1,\hat k)) 
		\right].
	\end{split}
\end{equation}
We find that, performing the multipole decomposition of \eqref{eq:c1regintermediate} 
we recover \eqref{eq:Ctail}, with $\kappa_\ell$, $\pi_\ell$ given in \eqref{eq:harmonicnumbers}.
More precisely, the first term in each line of \eqref{eq:harmonicnumbers} comes from the first line of \eqref{eq:c1regintermediate}, while the second one comes from the second line. 

Combining  \eqref{eq:Ctail} with the multipole decomposition of $\tilde{\mathcal{B}}_{1O}^{(h)}$ in \eqref{eq:B1OiB1Oh}, this provides an amplitude-based derivation of the multipole decomposition of 
\begin{equation}
	\tilde{W}_\text{tail,1}^{\mu\nu}
	=
	\pi G m_1 \omega_1 \, \tilde{\mathcal{A}}_0^{\mu\nu}
	+
	i G m_1 \omega_1 \log\frac{\omega_1^2}{\mu^2}\,\tilde{\mathcal{A}}_0^{\mu\nu}
	+
	\frac{i}{2}\,\tilde{c}^{\text{(reg)}\mu\nu}_{1}
\end{equation}
which is given by 
\begin{subequations}\label{eq:TAIL1}
	\begin{align}
		\mathrm{U}_{\ell m}^\text{tail,1} 
		&= 
		2iGm_1  \omega_1  \mathrm{U}_{\ell m}^\text{tree} \left(\log\frac{\omega_1}{\mu}-\kappa_\ell-\frac{i\pi}{2}\right)\!,
		\\
		\mathrm{V}_{\ell m}^\text{tail,1} 
		&= 
		2iGm_1 \omega_1  \mathrm{V}_{\ell m}^\text{tree} \left(\log\frac{\omega_1}{\mu}-\pi_\ell -\frac{i\pi}{2}\right)\!.
	\end{align}
\end{subequations}
Once $\tilde{W}_\text{tail,1}^{\mu\nu}$ is obtained in this way from the tree-level waveform, the full half-odd PN part of the one-loop waveform, $\tilde{W}_\text{tail,1}^{\mu\nu}+\tilde{W}_\text{tail,2}^{\mu\nu}$, is obtained by particle-interchange symmetry as discussed above.
Eq.~\eqref{eq:TAIL1} is of course the probe limit of the celebrated tail formula,
\begin{subequations}\label{eq:TAIL}
	\begin{align}
		\mathrm{U}_{\ell m}^\text{tail} 
		&= 
		2iGE  \Omega  \, \mathrm{U}_{\ell m}^\text{tree} \left(\log\frac{\Omega}{\mu}-\kappa_\ell-\frac{i\pi}{2}\right)\!,
	 	\\
		\mathrm{V}_{\ell m}^\text{tail} 
		&= 
		2iGE \Omega \, \mathrm{V}_{\ell m}^\text{tree} \left(\log\frac{\Omega}{\mu}-\pi_\ell -\frac{i\pi}{2}\right)\!,
	\end{align}
\end{subequations}
where $E$ and $\Omega$ are defined in the center of mass, and we used \eqref{eq:EOmega}. However, the object $\tilde{W}_\text{tail}^{\mu\nu}$ obtained by recombining the multipoles in \eqref{eq:TAIL} does \emph{not} coincide with $\tilde{W}_\text{tail,1}^{\mu\nu}+\tilde{W}_\text{tail,2}^{\mu\nu}$ evaluated in that frame, and more non-linear combinations of tree-level multipoles are needed in order to reconstruct the full half-odd PN part of the waveform (see \cite{Blanchet:2013haa,Georgoudis:2024pdz,Bini:2024rsy} and references therein for more details). Conversely, the procedure discussed above bypasses the need for these terms by combining particle interchange symmetry and the boost connecting the rest frame with the center-of-mass frame.

\subsection*{Acknowledgments}
We would like to thank Andrea Cipriani, Stefano Foffa, Francesco Fucito, Alessandro Georgoudis, Francisco Morales and Riccardo Sturani for useful discussions.
R.~R. is partially supported by the UK EPSRC grant ``CFT and Gravity: Heavy States and Black Holes'' EP/W019663/1 and the STFC grant ``Amplitudes, Strings and Duality'', grant number ST/X00063X/1. No new data were generated or analyzed during this study.

\appendix

\section{Physical polarizations and transformation rules}
\label{app:Polarizations}

In this appendix, we recall some general properties of physical polarizations and their transformation laws under the Lorentz group and under reparametrizations. We then provide some specific expressions of little group phases relevant to the transformations discussed in the text.
See e.g.~\cite{Cusin:2024git} for an assessment of the impact of such transformation properties on gravitational-wave emissions by astrophysical sources.

\subsection{Physical polarizations and little-group phase}
Let us consider a parametrization in terms of $\omega$, $\xi^{A}$ with $A=1,2$,
\begin{equation}\label{eq:anyparametrization}
k^\mu = \omega\,n^\mu(\xi) 
\end{equation}
for the null vector $k^\mu$.
A standard choice for this parametrization is
\begin{equation}\label{eq:parametrization}
	k^\mu = \omega \, n^\mu(\xi) \,, \qquad
	n^\mu(\xi) = (1,\sin\theta \cos\phi, \sin\theta \sin\phi, \cos\theta)
\end{equation}
with $\xi^A = (\theta, \phi)$ for $A=1,2$, but below we will consider a generic parametrization.

The coordinate tangent vectors
\begin{equation}
	e_A^\mu = \frac{\partial n^\mu}{\partial \xi^A}\,,
\end{equation}
which are transverse $e_A^\mu n_\mu = 0$,
give rise to the metric
\begin{equation}
	e_A^\mu e_{B\mu} = \gamma_{AB} \,.
\end{equation}
One can the introduce orthonormal polarizations by noting that $\gamma_{AB}$ can be decomposed in terms of a ``zweibein''
\begin{equation}\label{eq:zweibein}
	\gamma_{AB} = E\indices{_A^C} \delta_{CD} E\indices{_B^D}\,.
\end{equation}
Thus, letting
\begin{equation}\label{eq:defeps}
	\varepsilon_A^\mu = (E^{-1})\indices{_A^B}\,e_B^\mu\,,
\end{equation}
we obtain the desired orthonormal basis,
\begin{equation}
	\varepsilon_A^\mu \varepsilon^{\phantom{\mu}}_{B\mu} = 	\delta_{AB} \,.
\end{equation}
Note that the zweibein is only defined up to (local) rotations,
\begin{equation}\label{eq:localrot}
	E\indices{_A^B} \mapsto E\indices{_A^C} R\indices{^B_C}\,,\qquad
	(E^{-1})\indices{_A^B} \mapsto R\indices{_A^C} (E^{-1})\indices{_C^B} 
\end{equation}
with $R\indices{_A^B}$ an orthogonal matrix, $R\indices{_A^C} \delta_{CD}\,R\indices{_B^D}=\delta_{AB}$. These characterize different possible choices of polarization basis,
\begin{equation}
	\varepsilon_{A}^\mu \mapsto R\indices{_A^B}\, \varepsilon_B^\mu\,.
\end{equation}
The convenient choice of polarization ``vectors'' corresponding to the standard parametrization \eqref{eq:parametrization} is
\begin{equation}\label{eq:defeps1}
	\varepsilon_\theta^\mu (k) = \frac{\partial n^\mu}{\partial \theta}\,,
	\qquad
	\varepsilon_\phi^\mu(k) = \frac{1}{\sin\theta}\,\frac{\partial n^\mu}{\partial \phi}\,,
\end{equation}
so that $\varepsilon_A \cdot n = 0$ and $\varepsilon_A\cdot \varepsilon_B = \delta_{AB}$. One can also define polarization ``tensors'' according to 
\begin{equation}
	\varepsilon_{(+)}^{\mu\nu} = \frac{1}{\sqrt{2}}\,\left(
	\varepsilon_\theta^\mu \varepsilon_\theta^\nu - \varepsilon_\phi^\mu \varepsilon_\phi^\nu
	\right),
	\qquad
	\varepsilon_{(\times)}^{\mu\nu} = \frac{1}{\sqrt{2}}\,\left(
	\varepsilon_\theta^\mu \varepsilon_\phi^\nu + \varepsilon_\theta^\nu \varepsilon_\phi^\mu
	\right),
\end{equation}
which are transverse, traceless and orthonormal.  It  is also convenient to introduce the complex combinations
\begin{equation}\label{eq:defpols}
	\varepsilon_\pm^\mu = \frac{1}{\sqrt 2} \left( \varepsilon_\theta^\mu \pm i \varepsilon_\phi^\mu \right)\,,
	\qquad
	\varepsilon_{\pm}^{\mu\nu}
	= 
	\varepsilon_\pm^\mu \varepsilon_\pm^\nu = \frac{1}{\sqrt{2}}\left(
	\varepsilon_{(+)}^{\mu\nu} \pm  i \varepsilon_{(\times)}^{\mu\nu}
	\right).
\end{equation}
The latter can be easily  generalized to any integer spin $s$ by letting $\varepsilon_{\pm}^{\mu_1\cdots \mu_s} = \varepsilon_\pm^{\mu_1} \cdots \varepsilon_{\pm}^{\mu_s}$.

A Lorentz transformation maps the light-cone to itself and thus induces a mapping
\begin{equation}
	\Lambda\indices{^\mu_\nu} \,n^\nu(\xi) = \Omega(\xi,\Lambda)\,n^\mu(\xi')\,,
\end{equation}
with $\xi'^A=\xi'^A(\xi,\Lambda)$.
Taking a derivative of this relation with respect to $\xi^A$, we find
\begin{equation}\label{eq:transforme}
	\Lambda\indices{^\mu_\nu} \,e^\nu_A(\xi) = \Omega(\xi,\Lambda) \, \frac{\partial \xi'^B}{\partial \xi^A}\,e_B^\mu(\xi') + \left[E^{-1}(\xi)\right]\indices{_A^B}\, \frac{\partial \Omega(\xi,\Lambda)}{\partial \xi^B}\,n^\mu(\xi')
\end{equation}
and ``squaring'' this relation shows that the mapping $\xi'^A(\xi,\Lambda)$ is a conformal transformation for the metric $\gamma_{AB}$,
\begin{equation}\label{eq:conformal}
	\gamma_{AB}(\xi) = \Omega(\xi,\Lambda)^2\,\frac{\partial \xi'^C}{\partial \xi^A}\,\gamma_{CD}(\xi')\,\frac{\partial \xi'^{C}}{\partial \xi^B}\,.
\end{equation}
Introducing the decomposition \eqref{eq:zweibein} in terms of the zweibein, we can use the relation \eqref{eq:conformal} to check that the matrix defined by
\begin{equation}\label{eq:odef}
	O(\xi,\Lambda) \indices{_A^B}= \Omega(\xi,\Lambda)\, [E^{-1}(\xi)]\indices{_A^C}\,\frac{\partial \xi'^D}{\partial \xi^C}\,E(\xi')\indices{_D^B}
\end{equation}
is orthogonal,
\begin{equation}
	O(\xi,\Lambda)\indices{_A^B}\, \delta_{BD}\,O(\xi,\Lambda)\indices{_C^D} = \delta_{AC}\,.
\end{equation}
Similarly, one can re-express the transformation law \eqref{eq:transforme} in terms of $\varepsilon_A^\mu$ using its definition, and one finds
\begin{equation}\label{eq:transformeps}
	\Lambda\indices{^\mu_\nu} \,\varepsilon^\nu_A(\xi) = O(\xi,\Lambda) \indices{_A^B}\,\varepsilon_B^\mu(\xi') + \frac{\partial \Omega(\xi,\Lambda)}{\partial \xi^A}\,n^\mu(\xi')\,.
\end{equation}
A convenient way of calculating the rotation matrix is to contract both sides of \eqref{eq:transformeps} with the ``new'' polarization, $\varepsilon_{B\mu}(\xi')$ (note that the indices of $O_{AB}$ can be raised and lowered freely)
\begin{equation}\label{eq:ocalc}
	O(\xi,\Lambda)_{AB} = \varepsilon_{B\mu}(\xi')\, \Lambda\indices{^\mu_\nu} \,\varepsilon^\nu_A(\xi) \,.
\end{equation}
Under the local rotations \eqref{eq:localrot}, the little-group matrix transforms as follows,
\begin{equation}\label{eq:ambiguity}
	O(\xi,\Lambda)_{AB}  \mapsto R(\xi)\indices{_A^C} O(\xi,\Lambda)_{CD} R(\xi')\indices{_B^D}\,,
\end{equation}
as one easily deduces from \eqref{eq:odef} or \eqref{eq:ocalc}.

Let us now consider the amplitude $\mathcal{A}_\mu(k,p)$ 
for the emission of
a massless spin-one state with momentum $k^\mu$ (we use $p$ as a shorthand for a generic collection of any number of additional momenta)
and define its contraction with the polarization as follows,
\begin{equation}
	\mathcal{A}_A(k,p) = \varepsilon_A^\mu(\xi) \mathcal{A}_\mu (\omega\, n(\xi), p)\,.
\end{equation} 
Using the transformation rule \eqref{eq:transformeps}, the Lorentz covariance of $\mathcal{A}_\mu$, 
\begin{equation}
	\mathcal{A}_{\mu}(k',p') = \Lambda\indices{_{\mu}^{\nu}}\,\mathcal{A}_{\nu}(k,p)\,, \quad p'^\mu = \Lambda\indices{^\mu_\nu}\, p^\nu \,, \quad k'^\mu = \Lambda\indices{^\mu_\nu}\, k^\nu\,,
\end{equation}
(we recall that $(\Lambda^{-1})\indices{^\mu_\nu}=\Lambda\indices{_\nu^\mu}$ for Lorentz transformations)
and gauge invariance, 
\begin{equation}
k^\mu \mathcal{A}_\mu(k,p)=0\,,
\end{equation}
we thus obtain the  transformation law
\begin{equation}
	O(\xi,\Lambda)\indices{_A^B} \mathcal{A}_B(k',p') = \mathcal{A}_A(k,p)\,.
\end{equation}
Since $O$ is a 2D rotation matrix, it can be parametrized in the standard way
\begin{equation}\label{eq:standardform}
	O(\xi,\Lambda) = \left(\begin{matrix}
		\cos\Theta(\xi,\Lambda) & \sin\Theta(\xi,\Lambda)\\
		-\sin\Theta(\xi,\Lambda) & \cos\Theta(\xi,\Lambda)
	\end{matrix}\right)\,.
\end{equation}
Therefore, letting 
\begin{equation}
	\varepsilon^\mu_{\pm} = \frac{1}{\sqrt{2}}\left(\varepsilon_1^\mu \pm i \varepsilon_2^\mu\right),\qquad
	\mathcal{A}_{\pm}(k,p) = \varepsilon^{\ast\mu}_{\pm}(\xi) \mathcal{A}_{\mu}(\omega\,n(\xi),p) \,,
\end{equation}
we obtain
\begin{equation}
	e^{\pm i \Theta(\xi,\Lambda)} \mathcal{A}_{\pm}(k',p') = \mathcal{A}_{\pm}(k,p)
\end{equation}
for the transformation of an amplitude with definite helicity.

For an amplitude involving the emission of a massless spin-$s$ particle with $\pm$ helicity, we define
\begin{equation}\label{eq:defApm}
	\mathcal{A}_{\pm}(k,p) = \varepsilon_{\pm}^{\ast\mu_1}\cdots \varepsilon_{\pm}^{\ast\mu_s}(k) \mathcal{A}_{\mu_1\cdots \mu_s}(k,p)\,,
\end{equation}
where $\mathcal{A}_{\mu_1\cdots \mu_s}(k,p)$ is a symmetric Lorentz tensor, so that when $k'^\mu = \Lambda\indices{^\mu_\nu}\,k^\nu$ (similarly for $p^\mu$),
\begin{equation}\label{eq:Lorentztensor}
	\mathcal{A}_{\mu_1\cdots \mu_s}(k',p') = \Lambda\indices{_{\mu_1}^{\nu_1}}\cdots \Lambda\indices{_{\mu_s}^{\nu_s}}\,\mathcal{A}_{\nu_1\cdots \nu_s}(k,p)\,.
\end{equation}
Moreover, $\mathcal{A}_{\mu_1\cdots \mu_s}$ is gauge invariant in the sense that
\begin{equation}\label{eq:transverse}
	k^{\mu_1}\mathcal{A}_{\mu_1\cdots \mu_s}(k,p)=0\,.
\end{equation}
Combining \eqref{eq:transformeps} and \eqref{eq:Lorentztensor} with \eqref{eq:defApm}, thanks to gauge invariance we finally obtain 
\begin{equation}\label{eq:finalspins}
	e^{\pm i s \Theta(k,\Lambda)} \mathcal{A}_{\pm}(k', p') = \mathcal{A}_{\pm}(k,p)\,.
\end{equation}

To summarize, \eqref{eq:odef} provides  a parametrization-independent construction of the little group \cite{Weinberg:1964ew} rotation $O(\xi,\Lambda)_{AB}$, which one can easily calculate from \eqref{eq:ocalc}. One can then read off the phase $\Theta(\xi,\Lambda)$ from the representation \eqref{eq:standardform}. 
Adopting the same standard parametrization 
\begin{equation}\label{eq:standardformR}
	R(\xi) = \left(\begin{matrix}
		\cos\varphi(\xi) & \sin\varphi(\xi)\\
		-\sin\varphi(\xi) & \cos\varphi(\xi)
	\end{matrix}\right)
\end{equation}
also for the rotation matrix \eqref{eq:localrot} for possible changes of polarization basis, we see that the change \eqref{eq:ambiguity} in $O(\xi,\Lambda)$ translates into the following one in the phase $\Theta(\xi,\Lambda)$,
\begin{equation}
	\label{eq:ambiguityTheta}
	\Theta(\xi,\Lambda) \mapsto \Theta(\xi,\Lambda)+\varphi(\xi)-\varphi(\xi')\,.
\end{equation}

Choosing a different parametrization compared to \eqref{eq:anyparametrization}, 
\begin{equation}\label{eq:diffpar}
	k^\mu = \tilde \omega \, \tilde n^\mu (\tilde \xi)\,,
\end{equation}
with parameters $\tilde{\omega}$ and $\tilde \xi^A$ with $A=1,2$, will give rise to analogous quantities, in particular $\tilde O(\tilde \xi,\Lambda)_{AB}$ and $\tilde\Theta(\tilde\xi,\Lambda)$, which can in general be different even under the same Lorentz transformation. What is the relation, say, between $\tilde \Theta(\tilde \xi,\Lambda)$ and $\Theta(\xi,\Lambda)$? The equivalence of the two parametrizations demands that
\begin{equation}
	n^\mu ( \xi)=\sigma(\xi) \, \tilde n^\mu (\tilde \xi)\,,
\end{equation}
for a suitable function $\sigma(\xi)$.
Then, taking derivatives we find
\begin{equation}\label{eq:transepstilde}
	e_A^\mu(\xi)
	=
	\sigma(\xi)\,
	\frac{\partial\tilde\xi^B}{\partial\xi^A}\,\tilde{e}_B^\mu(\tilde\xi)
	+
	\frac{\partial\sigma(\xi)}{\partial \xi^A}
	\,\tilde{n}^\mu(\tilde\xi)
\end{equation}
and ``squaring'' this relation
\begin{equation}
	\gamma_{AB}(\xi)
	=
	\sigma(\xi)^2\,
	\frac{\partial\tilde\xi^C}{\partial\xi^A}\,\tilde{\gamma}_{CD}(\tilde\xi)\,\frac{\partial\tilde\xi^D}{\partial\xi^B}\,.
\end{equation}
Note that $\tilde{\gamma}_{AB}(\tilde{\xi})$ and $\gamma_{AB}(\xi)$ do not denote the same metric written in different coordinate systems in this case. Therefore, the relation between the zweibeins can be adjusted as follows
\begin{equation}
	E(\xi)\indices{_A^B} = \sigma(\xi) \,\frac{\partial\tilde\xi^C}{\partial \xi^A}\,\tilde{E}(\tilde\xi)\indices{_C^B} \,,
	\qquad
	[E(\xi)^{-1}]\indices{_A^B}
	=
	\frac{1}{\sigma(\xi)}\, 
	[\tilde E(\tilde\xi)^{-1}]\indices{_A^C} \,\frac{\partial \xi^B}{\partial \tilde\xi^C} 
\end{equation}
up to a \emph{suitable choice} of basis performed by using the rotation freedom in \eqref{eq:localrot}. Using the definition \eqref{eq:defeps} and the transformation law \eqref{eq:transepstilde}, we find that in this way the polarizations themselves coincide up to a ``pure gauge'' part,
\begin{equation}
	\varepsilon_A^\mu(\xi) = \frac{1}{\sigma(\xi)}\,[\tilde E(\tilde\xi)^{-1}]\indices{_A^C} \,\frac{\partial \xi^B}{\partial \tilde\xi^C} \,e_B^\mu(\xi)
	=
	\tilde{\varepsilon}_A^\mu(\tilde\xi)
	+
	[E(\xi)^{-1}]\indices{_A^B}\,\frac{\partial\sigma(\xi)}{\partial \xi^B}
	\,\tilde{n}^\mu(\tilde\xi)
\end{equation}
so that
\begin{equation}\label{eq:adjusted}
	\varepsilon_A^\mu(\xi) \tilde\varepsilon_{B\mu}(\tilde\xi) = \delta_{AB}\,.
\end{equation}
From the definition \eqref{eq:odef} of $O$, we then see that
\begin{equation}\label{eq:olink}
	O(\xi,\Lambda) \indices{_A^B}= \frac{\Omega(\xi,\Lambda)\,\sigma(\xi')}{\sigma(\xi)}\, [\tilde E^{-1}(\tilde\xi)]\indices{_A^C} \frac{\partial \xi^D}{\partial \tilde \xi^C} \,\frac{\partial \xi'^E}{\partial \xi^D}\,\frac{\partial \tilde\xi'^F}{\partial \xi'^E}\,\tilde E(\tilde\xi')\indices{_F^B}
\end{equation}
so that thanks to the chain rule
\begin{equation}
	O(\xi,\Lambda) \indices{_A^B}= \frac{\Omega(\xi,\Lambda)\,\sigma(\xi')}{\sigma(\xi)}\, [\tilde E^{-1}(\tilde \xi)]\indices{_A^C} \frac{\partial \tilde\xi'^D}{\partial \tilde \xi^C} \,\tilde E(\tilde\xi')\indices{_D^B}\,,
\end{equation}
and finally, using again the definition \eqref{eq:odef}, now with the new parametrization, we obtain
\begin{equation}
	O(\xi,\Lambda) \indices{_A^B}= \frac{\Omega(\xi,\Lambda)\,\sigma(\xi')}{\tilde{\Omega}(\tilde\xi,\Lambda)\,\sigma(\xi)}\, 
	\tilde O(\tilde \xi,\Lambda) \indices{_A^B}\,.
\end{equation}
By consistency, since both $O$ and $\tilde O$ are unitary matrices, the prefactor must be $\pm1$.  This shows that, for any two given parametrization, there exists a  choice of polarization basis for which the Wigner rotations and phase take the same form. Conversely, if one does not adjust the basis in this way, then the relation between polarizations will be
\begin{equation}
	\varepsilon_A^\mu(\xi) 
	=
	R(\xi)\indices{^B_A}\,
	\tilde{\varepsilon}_B^\mu(\tilde\xi)
	+
	[E(\xi)^{-1}]\indices{_A^B}\,\frac{\partial\sigma(\xi)}{\partial \xi^B}
	\,\tilde{n}^\mu(\tilde\xi)
\end{equation}
and instead of \eqref{eq:adjusted} we have
\begin{equation}\label{eq:notadjusted}
	\varepsilon_B^\mu(\xi) \tilde\varepsilon_{A\mu}(\tilde\xi) = R(\xi)_{AB}\,.
\end{equation}
The two rotations and phases will be related by \eqref{eq:ambiguity}, \eqref{eq:ambiguityTheta}, so that
\begin{equation}\label{eq:newrottilde}
	\tilde{O}(\xi,\Lambda)_{AB}
	=
	R(\xi)\indices{_A^C} \, O(\xi,\Lambda)_{CD}  \, R(\xi')\indices{_B^D}
\end{equation}
and
\begin{equation}\label{eq:newThetatilde}
	\tilde{\Theta}(\tilde \xi, \Lambda) = \Theta(\xi, \Lambda) + \varphi(\xi) - \varphi(\xi')\,.
\end{equation}
In other words, changing parametrization will generally induce a new choice of polarizations that differs by a local rotation from the original one and consequently a change in the little group phase that can be calculated from \eqref{eq:notadjusted} via \eqref{eq:newrottilde}, \eqref{eq:newThetatilde}.

\subsection{Specific boosts and rotations}

Following the conventions in \cite{Georgoudis:2024pdz,Bini:2024ijq}, we align the particles' momenta along $y$, while adopting the standard parametrization \eqref{eq:parametrization}. So, to go from the center-of-mass frame to the rest frame of particle 1, we need to perform the following boost,
\begin{equation}\label{eq:boosty}
	B_y = \left(\begin{matrix}
		\frac{E_1}{m_1} & 0 &  -\frac{p}{m_1} & 0\\
		0 & 1 & 0 & 0\\
		-\frac{p}{m_1} & 0 & \frac{E_1}{m_1} & 0\\
		0 & 0 & 0 & 1\\
	\end{matrix}\right)
\end{equation}
which maps $-p_1^\mu=(E_1, 0, p, 0)$ to $(m_1, 0,0,0)$.
Constructing the little group transformation associated to $\Lambda$ and reading off the rotation angle $\Theta$, one finds that it is nontrivial:
\begin{equation}\label{eq:phaseboosty}
	\Theta(\xi,B_y) =  
	\arctan\left[
	(m_1+m_2\sigma) \sin\theta - m_2 p_\infty\, \sin\phi ,
	m_2 p_\infty \cos\theta \cos\phi\,
	\right]
	\,.
\end{equation}

Instead, one finds a trivial little-group phase for boosts along the $z$ axis,
\begin{equation}
	B_z = \left(\begin{matrix}
		\frac{E_1}{m_1} & 0 & 0 & -\frac{p}{m_1}\\
		0 & 1 & 0 & 0\\
		0 & 0 & 1 & 0\\
		-\frac{p}{m_1} & 0 & 0 & \frac{E_1}{m_1}
	\end{matrix}\right)
\end{equation}
which maps $-p_1^\mu=(E_1, 0, 0, p)$ to $(m_1, 0,0,0)$, 
\begin{equation}
	\Theta(\xi,B_z)=0\,.
\end{equation}
Of course, the reason for this different behavior is that the standard parametrization \eqref{eq:parametrization} assigns a privileged role to the $z$ axis. 
Suppose that, instead, we chose one privileging the $y$ axis,
\begin{equation}\label{eq:rotated}
	k^\mu = \omega \, \tilde{n}^\mu(\tilde\xi)\,,\qquad 
	\tilde{n}^\mu (\tilde\xi) = (1,\sin\tilde\theta\sin\tilde\phi, \cos\tilde\theta,\sin\tilde\theta\cos\tilde\phi)\,.
\end{equation}
Then the result must be that the phase is trivial for the $y$-boost, $\tilde{\Theta}(\tilde{\xi}, B_y)=0$. To see that this is the case, we can check that the phase associated to the rotation induced by the reparametrization according to \eqref{eq:notadjusted} is given by the standard form \eqref{eq:standardformR} with 
\begin{equation}
	\varphi(\xi) = \arctan\left( - \cos\phi\,,
	-\, \cos\theta \,\sin\phi\right)
\end{equation}
and that 
\begin{equation}
	\tilde{\Theta}(\tilde \xi, B_y)
	=
	\Theta(\xi,B_y) + \varphi(\xi) - \varphi(\xi') = 0
\end{equation}
as desired.

For the rotation by an angle $\phi_0$ around the $x$ axis,
\begin{equation}
	R_x(\phi_0) = \left(\begin{matrix}
		1 & 0 & 0 & 0\\
		0 & 1 & 0 & 0\\
		0 & 0 & \cos\phi_0 & -\sin\phi_0\\
		0 & 0 &\sin\phi_0 & \cos\phi_0 
	\end{matrix}\right),
\end{equation}
using the standard parametrization,
one also obtains a nontrivial little group phase
\begin{equation}\label{eq:lgphase-x}
	\Theta(\xi,R_x(\phi_0)) = \operatorname{arccot}\left(\cot\phi_0\,\frac{\sin\theta}{\cos\phi}-\cos\theta\,\tan\phi\right).
\end{equation}

Since
\begin{equation}
	B_y R_x(-\tfrac{\pi}{2}) = R_x(-\tfrac{\pi}{2}) B_z\,,
\end{equation}
one can check that the results above are consistent among each other because
\begin{equation}
	\Theta\left( R_x(-\tfrac{\pi}{2})(\xi) , B_y  \right) + \Theta\left( \xi, R_x(-\tfrac{\pi}{2}) \right)
	=
	\Theta\left( B_z(\xi) , R_x(-\tfrac{\pi}{2}) \right) + 0 \,,
\end{equation}
where we use the notation $\Lambda(\xi)=\xi'$ to specify which mapping is being used for the angular parameters.

\providecommand{\href}[2]{#2}\begingroup\raggedright\endgroup

\end{document}